# Linear Exponential Instability of the Hagen-Poiseuille Flow with Respect to Synchronous Bi-Periodic Disturbances

Sergey G. Chefranov [1)], Alexander G. Chefranov [2)]

1) A.M. Obukhov Institute of Atmospheric Physics, Russian Academy of Sciences; Rheumatology Institute, Russian Academy of Medical Sciences, Moscow, Russia; e-mail: schefranov@mail.ru

2) Eastern-Mediterranean University, Famagusta, North Cyprus; Institute of Technology, Southern Federal University, Taganrog, Russia; e-mail: Alexander.chefranov@emu.edu.tr

For Gagen-Poiseuille flow, we show that exponential instability (to extremely small, axially symmetric disturbances represented by Galerkin's approximation) is possible only if there exists bi-periodic variability of the disturbances along the pipe axis when the threshold Reynolds number $\mathrm{Re}_{th}(p)$ depends on the ratio $p$ of two longitudinal periods. Absolute minimum $\mathrm{Re}_{th}(p) \approx 448$ (for $p \approx 1.53..$) is obtained that corresponds to the observed conditions of transition from the laminar resistance law to the turbulent one and Tollmien-Schlichting waves exciting in the boundary layer.

1. Fundamental and applied problem of understanding of the mechanism of turbulence development for the Gagen-Poiseuille (GP) [1] flow remains mysterious because of the paradox of the linear stability of this flow to extremely small magnitude disturbances for any large Reynolds number values $\mathrm{Re}$ [1-4]. Up to now, to bypass this obvious paradoxical contradiction to experiment, a consensus is made to allow only non-linear mechanism of the GP flow instability to a disturbance with a sufficiently large finite magnitude [5-8]. This assumption is usual (see [3, 4]) based on the very special interpretation of the experiments in which the substantial increase of the threshold Reynolds number $\mathrm{Re}_{th}$ value (defining transition from the laminar to turbulent flow state) is achieved by increasing the smoothness of the streamlined pipe surface. In such an interpretation, only correlation between smoothness increasing and resulting decreasing of the initial disturbance magnitude is taken into account. Meanwhile, noted already by O. Reynolds [1] extremely high sensitivity of the value of $\mathrm{Re}_{th}$ to the initial disturbance does not exclude possibility of affecting on $\mathrm{Re}_{th}$ not only magnitude but frequency characteristics of the disturbances (resulting from non-ideal smoothness of the streamlined surface) as well. Actually, for example, in the experiment [9], it was found that under the fixed magnitude of artificially excited disturbances, GP flow instability develops only in the narrow disturbances frequency range.

Herein, we show that possibility of the linear absolute instability of the GP flow in the general case also [2] is defined by the value of the additional to $\mathrm{Re}$ control parameter $p$, which characterizes frequency properties of the disturbances and affects on the value of $\mathrm{Re}_{th}(p)$ independently from the magnitude of the initial disturbances. Such a parameter $p$ introducing is performed below on the base of pointed by O. Reynolds [1] (and by W. Heisenberg, see [4, 6]) dissipative GP flow instability mechanism[3] related with the action of molecular viscosity $\nu$ in the proximity of the solid boundary. According to [1], the mechanism manifests itself as a spontaneous one-stage appearance for $\mathrm{Re} > \mathrm{Re}_{th}$ of the vortexes having character size $l_{\nu}$, «..which does not grow any more contrary to expectations with the growth of the velocity magnitude [1]». As a result, in addition to the character disturbance scale related to the pipe diameter, $2R$, due to near-boundary vortex-generating viscosity action, there exists a new additional scale $l_{\nu}$ ,which is proportional to the size of some part (which is really activated only for $\nu \neq 0$) of the pipe boundary. It, together with $R$, can define frequency parameters of the initial disturbances, e.g., the longitudinal along the pipe axis (axis $z$) spatial periods. The ratio of the periods, $p = \frac{l_{\nu}}{2R}$, as it is shown below, is a new additional to $Re$ parameter defining the GP flow instability threshold to the conditionally periodic on $z$ extremely small vortex disturbances. Such representation of the disturbances structure meets in observed conditionally periodic Tollmien-Schlichting (TS) waves development of which (caused also by near-boundary action of the molecular viscosity) precedes explosive emerging of the turbulence in the near-boundary layer [17-19]. Moreover, already in [2, 20], it was noted that usually considered in the linear stability theory strictly periodic on $z$ structures of the disturbances field obviously do not match those observed in the experiments. We show further, that the assumption of the strict longitudinal (by $z$) periodicity of the disturbances may result in the paradoxical conclusion on linear stability of the GP flow for any $Re$ with $\mathrm{Re}_{th} \to \infty$. The same time, with the simplest Galerkin's approximation[4], we get the finite value of the absolute minimum $\mathrm{Re}_{th} \approx 448$ (for $p \approx 1.53..$), that is the observed characteristic threshold value Re for transition from the laminar resistance law to the turbulent one [2, 21] and for the conditions of the TS waves exciting [19]. Qualitatively the proposed theory conclusions agree with the experimental data for the flows in a pipe [22-24]; also, we have conducted comparison of our conclusions with conclusions of the stability theory (Schlichting and Lin's), and with experimental data on the stability of laminar near-boundary layer [25].

---

[1] GP flow is a laminar, stationary flow of the uniform viscous fluid along the static, direct and the length unbounded pipe with the round the same along the pipe axis crosscut.

[2] For special GP flow modifications (in the pipe with elliptic crosscut [10], or in the rotating pipe [11], or in the case of the flow with the finite size particles transported by it[12]), there initially exists natural additional to $\mathrm{Re}$ control parameter, and the paradox of linear stability is absent.

2. Let's consider the well known (see [4]) representation of the GP flow in the cylindrical coordinate system $(z,r,\varphi)$: $V_{0r} = V_{0\varphi} = 0, V_{0z} = V_{\max}(1-\frac{r^2}{R^2})$, where $V_{\max} = \frac{R^2}{4\rho\nu}\frac{\partial p_0}{\partial z}$, fluid density $\rho = const$, $\frac{\partial p_0}{\partial z}$ is a constant value of the pressure $p_0$ gradient along the axis of the pipe with radius $R$, and $\nu$ is the fluid kinematic viscosity coefficient.

In the axially symmetric case (i.e. for extremely small disturbances which are not depending on $\varphi$), linear instability of the GP flow may be defined only by the value of the tangential component of the disturbance velocity $V_\varphi$, which before coming to the non-linear evolution stage does not depend on other disturbance fields and satisfies the following equation:

$$\frac{\partial V_\varphi}{\partial t} + V_{0z}(r)\frac{\partial V_\varphi}{\partial z} = \nu(\Delta V_\varphi - \frac{V_\varphi}{r^2}) , \tag{1}$$

where $\Delta$ is the Laplace operator. It follows from (1) that exponential growth of $V_\varphi$ with time, not depending on the reference system choice (i.e. absolute instability of the flow [2, 4]), evidently is not realizable as for $\nu = 0$ so for the case of strictly periodic or localized by $z$ functions $V_\varphi$ in (1). It is easy to verify the latter, if to multiply (1) by $V_\varphi$ and integrate the both sides of (1) by $z$ in the limits of the period of the function $V_\varphi$ (or from $-\infty$ to $+\infty$ in the case of the localized by $z$ function $V_\varphi$), when the contribution of the second term in the left hand side of (1) is zero.

---

[3] Such a mechanism is realized in the systems having disturbances with the negative energy [13-16] as , for the threshold generation of the vortexes (rotons) in the stream of the superfluid helium in capillary [13].

[4] Galerkin approximation is used to represent disturbance dependency not only on radial but on the longitudinal coordinate z as well when for two synchronous radial modes having two different variability periods on z the order of the exponential temporal evolution is one and the same. The latter causes necessity of the use of Galerkin approximatiom for describing disturbance dependency on z contrary to the conventional problem stating [2, 3] in the linear theory of the GP flow stability where independent modes are considered – each with its own order of the exponential temporal evolution. Appendix B contains detailed example illustrating distinctions of our linear stability problem stating from the conventional one (see, e.g. [2, 3]).

Thus far, only for conditionally periodic by $z$ (see [2]) functions and only for $\nu \neq 0$ in (1) it is possible to expect realization of the absolute dissipative instability of the GP flow for above the threshold Reynolds numbers $Re > \mathrm{Re}_{th}$, where $Re = \frac{V_{\max} R}{\nu}$.

Let's find solution of (1) in Galerkin's approximation (see [3]), describing dependence of $V_\varphi$ on $r$ as follows

$$V_\varphi = V_{\max} \sum_{n=1}^{N} A_n(z,t) J_1(j_{1,n}\frac{r}{R}) , \tag{2}$$

where $V_\varphi$ in (2) for any $A_n$ meets necessary boundary conditions on $r$ ($V_\varphi < \infty$ for $r = 0$, and $V_\varphi = 0$ for $r = R$), because $J_1$ is the first-order Bessel function, and $\gamma_{1,n}$ are its zeroes, i.e. $J_1(\gamma_{1,n}) = 0$ for any $n = 1,2,..$ For coefficients $A_n$, from (1), (2), we get the following system of equations in the dimensionless form:

$$\frac{\partial A_m}{\partial \tau} + j_{1,m}^2 A_m - \frac{\partial^2 A_m}{\partial x^2} + \mathrm{Re}\sum_{n=1}^{N} P_{nm}\frac{\partial A_n}{\partial x} = 0, \tag{3}$$

where $m = 1,2,..,N, \tau = \frac{t\nu}{R^2}, x = \frac{z}{R}$. In (3), constant coefficients, $P_{nm}$, are of the form

$$P_{nm} = \frac{2}{J_2^2(j_{1,m})}\int_0^1 dy\, y(1-y^2) J_1(j_{1,n}y) J_1(j_{1,m}y) , \tag{4}$$

where $J_2$ is the second-order Bessel function, and the linear on $y$ term under the integral yields in $P_{nm}$ the contribution in the form of the unity matrix $\delta_{nm} = \begin{cases} 1, n = m \\ 0, n \neq m \end{cases}$. For $N = 1$ in (3), the last term may be excluded by the Galilean transformation, and hence for $N = 1$, there is no possibility for the absolute instability of the GP flow. In relation with this, we shall further consider (3) for the simplest non-trivial case, $N = 2$, that allows already to eliminate the GP flow linear stability paradox, and leads to the conclusions quantitatively agreeing with the experimental data (see section 4) [22, 24].

3. Observed in the experiment field structures do not agree with the strictly periodic along the pipe axis disturbances (see herein above, and [2, 20]). Moreover, in [20], it is noted that different radial modes (defining dependency on the radial coordinate) correspond to different variability periods along the pipe axis. The latter can be modeled with the use in (3) of different periods along the pipe axis for the modes with different index value m. Such a requirement corresponds to introducing for each of these modes of its own, independent from other modes, periodical boundary condition. Hence, it is necessary to use the Galerkin's approximation for a dependency of the solution of (3) on x relevant to the mentioned above boundary conditions. Let, thus far, in (3), for example for $N=2$, magnitudes $A_1$ and $A_2$ have the form of the running waves with different periods along the pipe axis:

$$A_1 = A_{10}e^{\lambda\tau+i2\pi\alpha x},\ A_2 = A_{20}e^{\lambda\tau+i2\pi\beta x}, \tag{5}$$

where $A_{10}$ and $A_{20}$ are constants, and the additional to Re control parameter is defined as $p=\frac{\alpha}{\beta}$ for all $\alpha$ and $\beta$, when, in particular, $p=\frac{\alpha}{\beta}=\frac{l_\nu}{2R}$. In Galerkin's approximation (for the solution of the system (3) depending on $x$ in the form (5) for $\alpha \neq \beta$ generally), we get a uniform system for $A_{10}$ and $A_{20}$, condition of solvability of which for the finite $A_{10}$ and $A_{20}$ allows to define the value of the exponent $\lambda=\lambda_1+i\lambda_2$ depending on dimensionless parameters $\mathrm{Re}$, $p$, and $\beta$ (see (A.1) and (A.2) in Appendix).

Vortex structure of the considered in (2), (5) disturbances is related to non-zero components (longitudinal and radial) of the vorticity field, and zero value of the integral value of the Lamb's vortex momentum density ,which for the main undisturbed GP flow has longitudinal component only equal to $\frac{\rho V_{\max}}{2}$. This structure has finite value of the integral helicity for the resulting flow which is a linear superposition of the main GP flow (having non-zero the longitudinal component of the velocity field and the azimuthal component of the vorticity only) and of the disturbance field.

Usage according to [20] of the form (5) to get solution of the system (3) just in the Galerkin's approximation (instead of usually applied exact solution of the system (3) for the radial modes having one and the same longitudinal period) may be also interpreted as a result of the possible modeling of the always existing stochastic heterogeneities of the pipe internal surface.

Linear exponential instability condition with $\lambda_1>0$ in (5) has the form of (A.3). For $\mathrm{Re}>>1$, it may be reduced to more simple (A.4).Then for $\beta$ from (A.5), minimizing $\mathrm{Re}_{th}$ in (A.4), it defines the following linear exponential instability condition for GP flow (providing a lower boundary for an exact value of $\mathrm{Re}_{th}$, determined from (A.3)):

$$\mathrm{Re} > \mathrm{Re}_{th} = \frac{\pi^2(1-p)^2 F^{1/2}}{P_{12}P_{21}p^2|S|}, \tag{6}$$

where $S=\sin\pi p\sin\frac{\pi}{p}\sin\pi(\frac{1}{p}+p)$, $B=\frac{S}{|S|}(pP_{11}-P_{22})$,

$F=(\gamma_{1,2}^2+\gamma_{1,1}^2)(1+p^2)A^2+(\gamma_{1,2}^2-\gamma_{1,1}^2)(1-p^2)B^2+2AB(\gamma_{1,2}^2-p^2\gamma_{1,1}^2)$,

$A^2=B^2-\frac{4S P_{12}P_{21}p^2 ctg\pi(p+\frac{1}{p})}{\pi^2(1-p)^2}$ for $P_{11},P_{22,}P_{12},P_{21}$ from (4) , for arbitrary value of $p$ and for $A^2>0$.

Only for $A^2 < 0$, it is necessary to use instead of (A.4) and (6) the general condition (A.3), for which $\lambda_1 > 0$ in (5).

In (6), infinite value $\mathrm{Re}_{th} \to \infty$ takes place only for those finite values of $p$ for which $S = 0$, i.e. for $p = p_k = k, p = p_{\frac{1}{k}} = \frac{1}{k}$, and $p = p_{\sqrt{k+1}} = \frac{k+1 \pm \sqrt{(k+1)^2 - 4}}{2}$ for all integer $k = 1,2,...$ For $p$, being inside the intervals of changing of $p$ between any two consecutive values of $p_k, p_{\frac{1}{k}}, p_{\sqrt{k+1}}$, value of $\mathrm{Re}_{th}$ in (6) is a function of $p$ having one local minimum inside each of the pointed out intervals (see Fig. 1,b). So far, the absolute minimum value, $\widetilde{\mathrm{Re}}_{th}^{\min} \approx 442$, in (6) is reached for $p \approx 1.53..$, close to the "golden" ratio $p_g = \frac{1+\sqrt{5}}{2} \approx 1.618..$ (i.e. the limit of the infinite sequence of ratios of two consequent Fibonacci numbers 1, 2, 3, 5, 8, 13, 21, ..). For the same $p$, from the exact condition (A.3), we get close to it absolute minimum value $\widetilde{\mathrm{Re}}_{th}^{\min} \approx 448$ (see also Table in Appendix where conclusions based on (A.3) and (6) are compared).

4. Stated value $\widetilde{\mathrm{Re}}_{th}^{\min} \approx 448$ corresponds to the interval of $\mathrm{Re} \in 300 \div 500$ noted in the experimental observation of the threshold transition from the laminar resistance law (for the flow in the pipe) to the turbulent one [2, 21], and of the TS waves exciting in the near–wall region of the boundary layer [19]. Observed in [1] and other experiments (see references in [23, 24]) hyper-sensitivity of the value of $\mathrm{Re}_{th}$ to the initial disturbances, actually, corresponds to the obtained in (6) dependency of $\mathrm{Re}_{th}$ on $p$ when, for example, $\mathrm{Re}_{th}$ in (6) changes approximately 600 times with the change of $p$ from 0,1186 to 0,1118 only. Neighboring local minima of $\mathrm{Re}_{th}$ in (6) may also significantly differ each from the other, when for $p \approx 2.23$, we have in (6) $\mathrm{Re}_{th} \approx 1982$, and for $p \approx 3.86$, we already have $\mathrm{Re}_{th} \approx 84634$. Enlarged fragments of the neutral stability curve corresponding to the condition (6) (see Fig. 1, b), are represented in Fig. 1, a, as dependence of the value of 1/2p on Re. They are plotted on the taken from the paper [25] figure (see Fig.12 in [25]) representing theoretical (Lin's and Schlichting's) neutral stability curves and corresponding to them experimental data related to the determining of the instability threshold in the boundary layer.

Considered in (2), (5) conditionally periodic on $z$ structures of the initial disturbance field $V_\varphi$ qualitatively agree with observed in [23] wave-like changes of $V_\varphi$. The latter are especially clearly seen in the proximity of the turbulence decay threshold for $\mathrm{Re} \approx 1750$ when the ratio of visible in Fig. 5d) from [23] character periods with $p \approx \frac{8}{5}$ and $p \approx \frac{13}{8}$ is close to the pointed out above "golden" $p_g$. In Fig. 2a), taken from [24] (see Fig. 4 в [24]), it is giving the dependency on Re for the constant velocity of the rear front of the observed in the flow in the pipe [24] turbulent spot together with the experimental data [22] related to the given in Fig. 2a) interval of the values of Re (dark triangles for the rear front of the turbulent spot, and white triangles for the leading front). Also, results following from the non-linear theory [8] and for the phase velocity $V_\beta = -\frac{\lambda_2 V_{max}}{2\pi\beta \operatorname{Re}}$ (scaled by the mean flow GP velocity $V_m = \frac{V_{max}}{2}$) defined in (A.6) from the representation of $\lambda_2$ in (A.2) for the neutral curve (i.e. when $\lambda_1 = 0$) are presented there.

Hence, a theory based on the simplest Galerkin's approximation with $N = 2$ in (3) already yields agreeing with experiments [22, 24]. The same time, its worth to mention that linear exponential instability in (6) turns out to be possible also for rational $p$ not equal to $p = p_k$ or $p_{\frac{1}{k}}$, but for irrational $p = p_{\sqrt{k+1}}$, $k = 1,2,..$, vise versa, $\operatorname{Re}_{th} \to \infty$ in (6). It means that these conclusions based on the Galerkin's approximation with $N = 2$ may in this respect differ from the made above (in the section 2) general conclusions on the base of the original equation (1). Future GP flow stability analysis for $N > 2$ in (3) shall clarify the reason for such a discrepancy.

Obtained results allow to fill a gap in the non-linear theory [7, 8] in which up to now instead of the linear exponential instability it was necessary to consider a stage of the seed algebraic instability (where small initial disturbance may grow only locally in time tending to zero with $t \to \infty$).

Noted herein new mechanism of the linear instability of the GP flow (taking into account possibility of existence of the bi-periodic initial disturbances described by the complementing to the Reynolds number Re parameter p being the ratio of two longitudinal periods) may be useful for the flow in a duct of square cross-section [26], the plane Poiseuille and Couette flows stability analysis. For the latter flows, as for the GP flow, also currently there is no matching of the experimental data and conclusions of the linear stability theory taking into account only pure periodic disturbances. For example, in relation with noted in [27] linear destabilization effect for the plane Poiseuille flow (when "the region where solid-fluid coexist stretches towards the wall" [27, p. 33]), pointed out bi-periodicity of the initial disturbance can, as in [12] (see footnote 2 on the page 1), be defined by the complementing character scale of the solid units transported by the flow of the fluid.

We are grateful to S.I. Anisimov, E. A. Novikov and N.A. Inogamov for useful comments and interest to the work.

## *Appendix A*

1. From (3) and (5) for $N = 2$ in Galerkin's approximation (representing disturbance dependency on the longitudinal coordinate $z$ when generally $\alpha \neq \beta$ in (5)), one gets for $\lambda = \lambda_1 + i\lambda_2$:

$$\lambda_1 = -\gamma_{1,1}^2 - 4\pi^2\beta^2 p^2 - \frac{1}{2}\left(a_1 \pm \frac{1}{\sqrt{2}} D_1^{1/2}\right), \tag{A.1}$$

$$\lambda_2 = -2\pi\beta p P_{11} \operatorname{Re} - \frac{1}{2}\left(a_2 \pm \frac{1}{\sqrt{2}} D_2^{1/2}\right), \tag{A.2}$$

where $D_1 = d_0^{1/2} + l$, $D_2 = d_0^{1/2} - l$, $l = a_1^2 - a_2^2 + 4c_1 \operatorname{Re}_1^2$

$$a_1 = \gamma_{1,2}^2 - \gamma_{1,1}^2 + 4\pi^2\beta^2(1-p^2), \quad a_2 = 2\pi\beta \operatorname{Re}(P_{22} - pP_{11}),$$

$$\operatorname{Re}_1^2 = \frac{\operatorname{Re}^2 P_{21}P_{12}p^2\beta^2}{(1-p)^2}, \quad d_0 = l^2 + 4(a_1 a_2 - 2\operatorname{Re}_1^2 d_1)^2, \quad d_1 = -4S,$$

$c_1 = -4Sctg\pi(p+\frac{1}{p})$, and $S$ is defined in (6).

Condition $\lambda_1 > 0$ leads to the inequality

$$(a\mathrm{Re}+b)^2 > c + \frac{d}{\mathrm{Re}^2} \tag{A.3}$$

where $a = \frac{4P_{21}P_{12}p^2\beta^2 S}{(1-p)^2}$, $b = \pi\beta(P_{22} - pP_{11})a_1$,

$d = a_3^2(\gamma_{1,1}^2 + 4\pi^2\beta^2 p^2)(\gamma_{1,2}^2 + 4\pi^2\beta^2)$, $a_3 = \gamma_{1,2}^2 + \gamma_{1,1}^2 + 4\pi^2\beta^2(1+p^2)$,

$c = a_3^2\beta^2(\pi^2(P_{22} - pP_{11})^2 - 4\frac{P_{12}P_{21}p^2 Sctg\pi(p+\frac{1}{p})}{(1-p)^2})$.

2. In the limit of $\mathrm{Re} >> 1$, inequality (A.3) for $c > 0$ is reduced to the inequality

$$\mathrm{Re} > \mathrm{Re}_{th}(\beta) = \frac{\sqrt{c} - b\cdot\frac{S}{|S|}}{a} \ . \tag{A.4}$$

In (A.4), function $\mathrm{Re}_{th}(\beta)$ has the minimal value (given in (6)) when

$$\beta = \beta_0 = \frac{1}{2\pi}\left[\frac{A(\gamma_{1,2}^2 + \gamma_{1,1}^2) + B(\gamma_{1,2}^2 - \gamma_{1,1}^2)}{A(1+p^2) + B(1-p^2)}\right]^{1/2} , \tag{A.5}$$

where $A$ and $B$ are defined in the main text (see (6)) for $A^2 > 0$.

3. On the neutral curve with $\lambda_1 = 0$ (i.e. when equality takes place in (A.4)), the phase velocity $V_\beta / V_m$ is as follows

$$V_\beta / V_m = pP_{11} + P_{22} \pm (\frac{D_2^{1/2}}{2\sqrt{2}\pi\beta\,\mathrm{Re}})_{\beta=\beta_{1,2}} , \tag{A.6}$$

where $\beta = \beta_{1,2}(\mathrm{Re}, p)$ corresponds to the replacement of inequality by equality in (A.4). With such replacing of inequality by equality in (A.4), one gets quadratic equation with respect to $\beta$. Its solution is as follows

$$\beta = \beta_{1,2} = \frac{\mathrm{Re} \pm \sqrt{\mathrm{Re}^2 - \mathrm{Re}_{th}^2}}{2\pi^2\delta_1} \tag{A.7}$$

with $\mathrm{Re} \ge \mathrm{Re}_{th}$, $\mathrm{Re}_{th}$ is from (6), when $\beta_{1,2} = \beta_0$ for $\mathrm{Re} = \mathrm{Re}_{th}$, and $\delta_1 = \pi((p^2+1)\sqrt{1-4S\delta} - (1-p^2)\frac{Sb_1}{|Sb_1|})\frac{|b_1|(1-p)^2}{|S|\,p^2 P_{12}P_{21}}$,

$b_1 = P_{22} - pP_{11}, \delta = \frac{p^2 P_{21}P_{12}}{(1-p)^2\pi^2 b_1^2}ctg\pi(p+\frac{1}{p})$.

4.

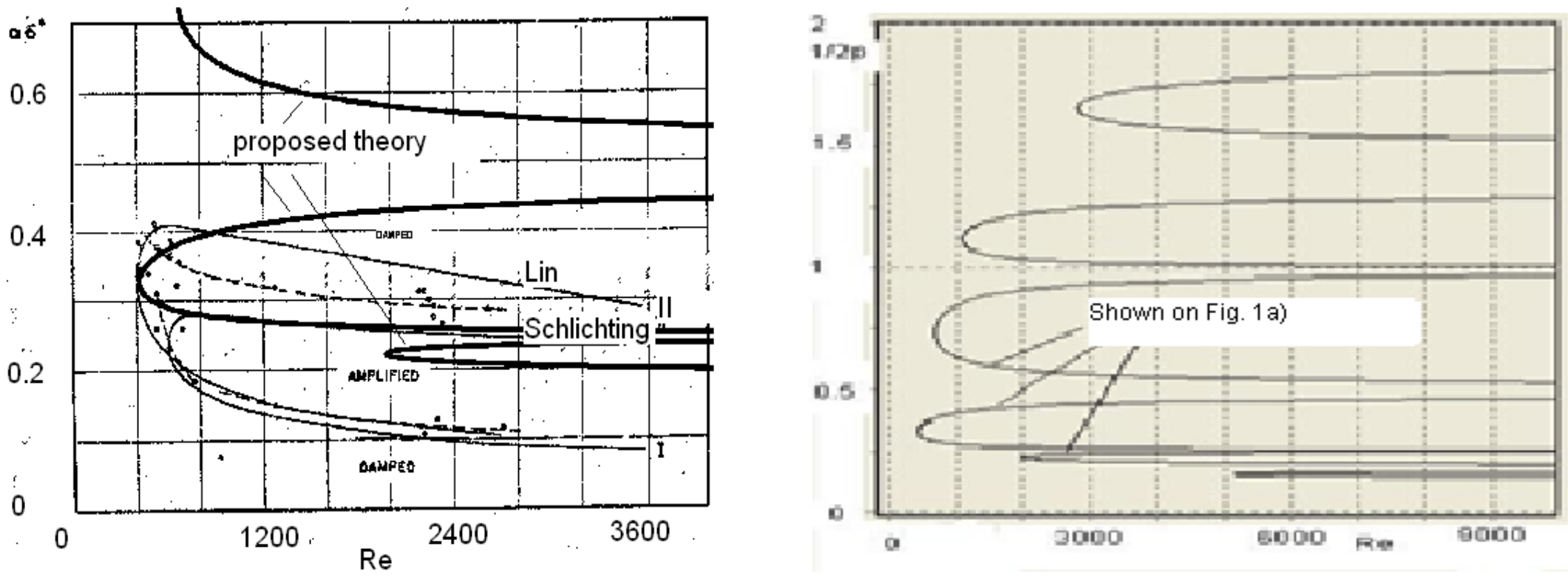

a) b)

Fig. 1. Family of the curves of the neutral stability (with $\lambda_1 = 0$) according to (6) and (A.5). In Fig.1a), a fragment from Fig.1b) is shown. In Fig. 1a), the upper curve (with two branches) corresponds to $\beta_0$=0.463 (for $p$=1.527), and the lower one to $\beta_0$=1.099 (for $p$=2.239) according to (A.5). Fig.1a) is overlapped with a figure from [25] (see Fig.12 in [25]) formally using *1/2p*= $\alpha\delta^*$. In [25], $\alpha$ is the disturbance wave number, and $\delta^*$ is the thickness of the boundary layer shift when streamlining a thin plate. In Fig. 1a), dots and dashed lines correspond to the experiment [25], and solid lines correspond to the theory of Schlichting (lower one with the Roman digit I), and Lin's (the upper one with the Roman digit II).

5. Table of values of $\mathrm{Re}_{th}$ and $\beta_0$ obtained by the formulae (A.3) and (6) for $p$ corresponding to the local minima of $\mathrm{Re}_{th}$ in the approximate formula (6)

| $p$ | $\beta_0$ (from (A.3)) | $\mathrm{Re}_{th}$ (from (A.3)) | $\beta_0$ (from (A.5)) | $\mathrm{Re}_{th}$ (from (6)) |
|---|---|---|---|---|
| 1,527 | 0,471 | 448,455 | 0,463 | 442,278 |
| 0,674 | 1,124 | 680,307 | 1,101 | 678,482 |
| 0,447 | 1,368 | 1095,455 | 1,358 | 1093,824 |
| 2,239 | 1,100 | 1983,171 | 1,099 | 1981,838 |
| 2,791 | 0,220 | 13095,398 | 0,219 | 13095,285 |
| 0,359 | 1,114 | 23816,499 | 1,114 | 23816,488 |

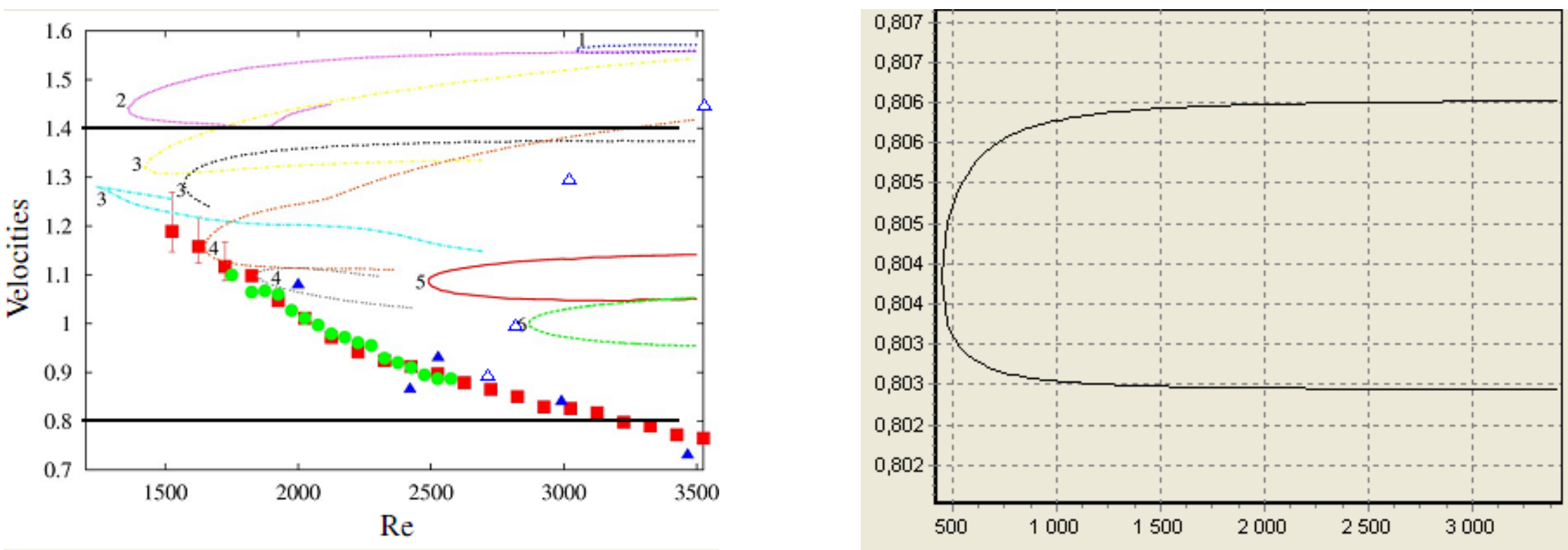

a) b)

Fig. 2, a) Results of the experiments from [22] ] (blue triangles are used for the turbulent spot rear front velocity, and white triangles reflect velocity of the leading front) and [24] (squares and rounds for the velocity of the rear front). Phase velocity of the wave solutions [8] (numbers 1-5 denote a degree of the azimuthal symmetry of the running wave). The velocities are scaled by the mean flow velocity in the pipe. Phase velocity (A.6) (upper "straight line" corresponds to the plus sign in (A.6), and the lower one corresponds to minus) for $p \approx 1.53$, $\beta$ =0,471 that corresponds to the absolute minimum $\mathrm{Re}_{th}$=448.5 according to (A.3). b) Enlarged representation of the lower "straight line" from Fig. 2, a) according to (A.6) (the upper branch in Fig. 2, b) corresponds to the plus sign in (A.7), and the lower one to minus.

## *Appendix B*

At first, let's consider conventional disturbance representation [2, 3]. Disturbance of the velocity field $V_\varphi$, described by (1), may be represented as the following two independent modes:

$$V_\varphi = V_{\varphi 1} = e^{\lambda_1 \tau + i2\pi\alpha x} \sum_{n=1}^{N} a_n J_1(\gamma_{1,n} \frac{r}{R}),$$
$$V_\varphi = V_{\varphi 2} = e^{\lambda_2 \tau + i2\pi\beta x} \sum_{n=1}^{N} b_n J_1(\gamma_{1,n} \frac{r}{R}) \qquad \text{(B.1)}$$

where $a_n, b_n$ are constants calculated by the use of the Galerkin's approximation for determining dependencies $V_{\varphi 1}$ and $V_{\varphi 2}$ on the radial coordinate $r$, $x = \frac{z}{R}$, $z$ is the longitudinal coordinate along the axis of the pipe of radius $R$, $J_1$ is the Bessel function of the first order ( $J_1(\gamma_{1,m}) = 0, m = 1,2,..$). For such disturbance representation, it is known [2, 3], that real parts of the orders of exponential temporal evolution $\lambda_1$ and $\lambda_2$, defined independently for each mode out of these two, are negative, and disturbances tend to zero when $\tau \to \infty$ for any Reynolds number. Obviously, it is true for the linear superposition of $V_{\varphi 1}$ and $V_{\varphi 2}$.

In our problem stating, for example, in the case of $N = 2$ in (3), we consider already quite different from (B.1) representation (5) of the disturbance $V_\varphi$ in the form of the set of two now synchronous modes, i.e. the modes have one and the same order of exponential temporal evolution $\lambda$ (and also different periods along the longitudinal axis $z$ for different radial modes):

$$V_\varphi = e^{\lambda \tau}(A_{10} e^{i2\pi\alpha x} J_1(\gamma_{1,1} \frac{r}{R}) + A_{20} e^{i2\pi\beta x} J_1(\gamma_{1,2} \frac{r}{R})). \qquad \text{(B.2)}$$

Here $\lambda$ may be defined only with the help of Galerkin's approximation for the used in (B.2) representation of the dependency of $V_\varphi$ on the longitudinal coordinate $z$, contrary to the exactly defined $\lambda_1$ and $\lambda_2$ in (B.1).

In Section 3 of the paper, we show that for the above-critical Reynolds number (6) real part of $\lambda$ in (B.2) can be already positive, i.e. GP flow can be exponentially unstable in the linear approximation.

## *References*

1. O. Reynolds, "An Experimental Investigation of the Circumstances which Determine whether the Motion of Water shall be Direct or Sinuous, and the Law of Resistance in Parallel Channels", Proc. Roy. Soc. (Lond.), 35, 84 (1883).
2. D. D.Joseph, Stability of fluid motion,Springer-Verlag,N.Y.1976.
3. P.G. Drazin, N.H. Reid, Hydrodynamic stability, Cambridge Univ. Press, Cambridge, England, 1981.
4. L.D. Landau, Ye. M. Lifshitz, Theoretical physics, v. 6, Hydrodynamics, Moscow, 2006
5. S. Grossman, Rev. Mod. Phys. 72, 603 (2000).
6. R. Fitzegerald, Physics Today, 57, 21 (2004).
7. H. Faisst, B. Eckhardt, Phys. Rev. Lett., 91, 224502 (2003).
8. H. Wedin, R. Kerswell, J. Fluid Mech. 508, 333 (2004).
9. J.A.Fox, M.Lessen, W.V.Bhat, Phys. Fluids, 11, 1 (1968).
10. R.R. Kerswell, A. Davey, J. Fluid Mech., 316, 307 (1996).
11. D.R. Barnes, R.R. Kerswell, J. Fluid Mech., 417, 103 (2000).
12. J.P. Matas, J.F. Morris, E. Guazzlli, Phys. Rev. Lett., 90, 014501 (2003).
13. L.D. Landau, JETP, 11, 592 (1941).
14. S.G. Chefranov, JETP Letters, 73, 312 (2001).
15. S. G. Chefranov, JETP, 126, 333 (2004).
16. S.G. Chefranov, Phys. Rev. Lett., 93, 254801 (2004).
17. A.S. Monin, A.M. Yaglom, Statistical hydromechanics, v. 1, Turbulence theory, St.-Petersburg, Hidrometeoizdat, 1992.
18. B.J. Cantwell, Ann. Rev. Fluid Mech. 13, 457 (1981).
19. Yu.S. Kachanov, V.V. Kozlov, V.Ya. Levchenko, Turbulence development in the boundary layer, Novosibirsk, 1982.
20. R.J. Leite, J. Fluid Mech. 5, 81 (1959).
21. S.J. Davies, C.M. White, Proc. Roy. Soc. (Lond. A), A 69, 92 (1928).
22. I.J. Wygnanski, F.H. Champagne, J. Fluid Mech., 59, 281 (1973).
23. J. Peixinho, T. Mullin, Phys. Rev. Lett., 96 , 094501 (2006).
24. B. Hof, C.W.H. van Doorne, J. Westerweel, F.T.M. Neiuwstadt, Phys. Rev. Lett., 95, 214502 (2005).
25. G.B. Schubauer, H.K. Skramstad, Laminar boundary layer oscillation and stability of laminar flow, NACA Rep. No. 909(1948).
26. D. Biau, H. S. Soueid, A. S. Bottaro, Transition to turbulence in duct flow, arXiv 1007.0081v1 [physics.flu-dyn] 1 Jul 2010.
27. M. Moyers-Gonzáleza, , T.I. Burgheleab, J. Mak, Linear Stability Analysis for Plane-Poiseuille Flow of an Elastoviscoplastic fluid with internal microstructure, arXiv: 1007.2392v1 [physics.flu-dyn] 14 Jul 2010.